# Contribution to the study of the flora in the central-west of Tunisia: landscape dynamics and evaluation of plant biodiversity of mountain Bouchebka

*Saadaoui Islem[1], Ilahi Hayet[1], Robin Bryant Christopher[2], and Rejeb Hichem[1]*

[1]Research Unit "Horticulture, Landscape and Environment",
University of Sousse, Higher Agonomic Institute of Chott Mariem,
BP 47, 4042 Sousse, Tunisia

[2]Department of Geography, Laboratory sustainable and dynamic territorial development,
University of Montreal / Faculty of Arts and Sciences,
C. P. 6128, Downtown succursale QC H3C 3J7, Montreal, Canada

**ABSTRACT:** The study was conducted during 2013 in Bouchebka, located in the central west of Tunisia. Such territory has a typical landscape of the transfrontier region. The series of the forest in Bouchebka is a part of the great mass of Aleppo pine. It is distinguished by the importance of the forest area which covers 92 % of the surface area (19,700 ha). The study attempts to inventory the natural vegetation and characterize ecological terms while highlighting the importance of environmental conditions. The method is based on a phytoecological analysis to quantify the floristic richness and diversity of the ecosystem in the forest of mountains in Bouchebka on the basis of floristic surveys and transects distributed in a stratified, systematic sampling in different vegetation formations that were previously distinguished. Statistical analyzes were performed using the Factorial Correspondence Analysis (FCA). The results show that the forest is composed primarily of the Aleppo pine trees, these forests are characterized by the abundance of young feet (10-25 cm diameter class). The ecosystem includes 12 families and 17 genera, 26 species. Thus the study has identified that the biological spectrum of the study area is characterized by a clear dominance of shrubs (41%) and chamaephytes (32 %). The distribution of plant species is influenced by ecological features of the region: the results show that 82% of species are drought tolerant which shows the arid environment. The region is also characterized by its windy side: 32% of species spread via anemochory. Factor analysis showed a pastoral aspect in the study area, with a presence of cultured human action exerted on the forest land species. Phytological spectrum indicates a predominance of woody species reflecting a territory dominated by open grassy areas, predominantly reflecting an arid climate.

## 1 INTRODUCTION

Tunisia, part of the Mediterranean basin, is one of the richest regions of the world in terms of biodiversity of flora and fauna [1]. Indeed, the Mediterranean countries hold almost 4.5% of the world's flora.

However, this plant heritage is threatened by degradation due to a combination of several natural factors (especially the recurrent droughts and arid climate) and anthropogenic, including overgrazing. This regressive dynamics of the natural vegetation has led several authors to sound the alarm about the risk, increasingly high, of attrition flora. Hence, the need to protect the natural vegetation cover, especially in arid areas, and assess the environmental impact and economic cost [2], [3].

Tunisian mountains cover about 2 million hectares, considering the terrain with elevations above 300 m. Areas of high potential contain vital resources. They are primary sources of water, agricultural land, grazing and forestry, which often enjoy good weather and are centers of biodiversity. The fragility of mountain ecosystems, interannual variability is associated with

# Contribution to the study of the flora in the central-west of Tunisia: landscape dynamics and evaluation of plant biodiversity of mountain Bouchebka

a Mediterranean climate. Proximity to the Sahara and the presence of a rural population in the mountains are the key factors in shaping landscapes observed. But also marked by strong natural and anthropogenic stresses generating competition, conflict and risk [4].

Of all the plant species that possess Tunisia, much of it is located in mountain ecosystems. Some are remarkable for their rarity, others by their abundance, some are considered for their economic interest, or simply scientific [5].

This article emanates from the study in the border zone: Bouchebka, the choice of this area is up to the need to conduct an impact on the vegetation inventory studies, monitoring and evaluation is an area border, landscape Bouchebka is a joint natural heritage between Tunisia and Algeria; a series of mountains belonging to the Tunisian dorsal and extending to the mountains of Tébessa Algeria.

This investigation attempts to characterize the floristic inventory of the landscape and vegetation of the area mountains of Bouchebka and diagnose the state of degradation taking into account the variables of natural environment and human interventions. The study distinguishes between plant formations and facies that constitute them. Indeed, vegetation identifying communities of plant species with faces more or less common features drawing landscapes.

## 2 METHOD

### 2.1 STUDY SITE

#### 2.1.1 FORESTS BOUCHEBKA: A COMPARTMENT OF THE ANTHROPIC ECOSYSTEM

The series of national forests of Bouchebka are part of the large solid mass of Alepo pines located on the plate ranging between Djebel Chambi and the mounts of Tebessa. These forests are located in the Gouvernate of Kasserine, Delegation of Feriana [6].

They cover 16991 ha (82.27%) of the total surface area of the forests in the delegation of Feriana.

Covering by the vegetation seems to depend primarily on the action of the Man livestock [7].

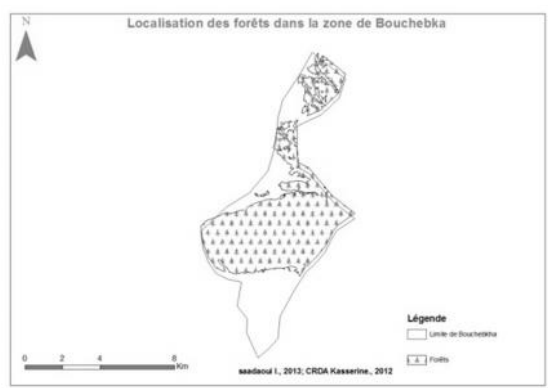

*Fig. 1. Map of the distribution of the forest area of Bouchebka*

### 2.2 GEOGRAPHIC LOCATION OF THE STUDY SITE

The series of national forests of Bouchebka fits entirely in the map to 1/50 000 Bouchebka (n) LXXXIII) between Lambert coordinates: X = 367-378 and Y = 207-215.

It is located in the Governorate of Kasserine, Delegation Feriana its limits, although sitting on the ground, are the following:

North: land of culture and Ain Amara forest which is separated from the first series of Dernaya by Ain Bou Deries track.

To the east: land of culture and forestry post Faider Remaïlia Sidi Baïssis.

South: land of culture Henchir el Goussa, the Ennafd el Bagrat until Henchir bel Houchet.

In the West: the land between the forest and the Algerian border cultures.

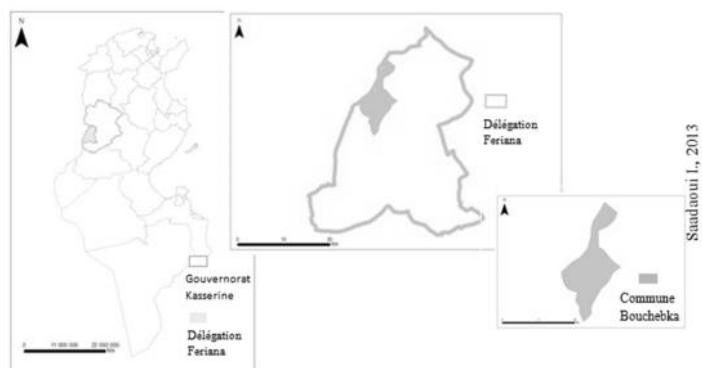

*Fig. 2. Geographic location of the municipality of Bouchebka*

## 2.3 METHOD

### 2.3.1 DETERMINING THE FLORISTIC COMPOSITION OF THE STUDY ENVIRONMENT

The botanical determination of the species is carried out directly on the ground in ecological stations of 10 x10 m ². When recognition of a species is not possible on the ground, a sample is kept in a herbarium and its identification is performed in the laboratory by means of the keys of botanical determination [8].

### 2.3.2 DESCRIPTIVE CHARACTERISTICS OF THE FLORA AND VEGETATION

The floristic list recorded on the ground (current list of the flora characteristic in the zone of the mountains), are supplemented by quantitative information and qualitative which will make it possible to characterize the flora and describe the vegetation. The quantitative information corresponds to various calculated indices of biodiversity, which characterize the flora of a site on a given date. The qualitative data are those of the auto ecology clean to the species, they inform on the distribution of the species according to their features of life.

The inventory carried out during the period of spring 2013 (END OF April and beginning of May), enabled us to collect 26 species different being in the area from study: the inventory covers the surface of one hectare pertaining to the forest post office of « Dernaya » ; it is a protected site (no human action grazing, deforestation ...).

Contribution to the study of the flora in the central-west of Tunisia: landscape dynamics and evaluation of plant biodiversity of mountain Bouchebka

Table 1. Number of the plants belonging to various plant species recorded

| Scientific name | code | Number of species (ni) |
|---|---|---|
| Pinus halepensis | Ph | 24 |
| Juniperus oxycedrus | Jo | 12 |
| Juniperus phoenicea | Jp | 9 |
| Phillyrea angustifolia | Pa | 2 |
| Scirpus sylvaticus | Ss | 12 |
| Stipa capillata | Sc | 145 |
| Stipa tenacissima | St | 12 |
| GLOBULARIA alypum | Ga | 116 |
| Artemisia campestris | Ac | 8 |
| Retama retam | Rr | 5 |
| Rosmarinus prostratus | Rp | 10 |
| Thymelaea hirsuta | Th | 15 |
| Diplotaxis simplex | Ds | 3 |
| **Scientific name** | **code** | **Number of species (ni)** |
| Diplotaxis harra | Dh | 3 |
| Centaurea calcitrapa | Cc | 9 |
| Artemisia vulgaris | Av | 6 |
| Rosmarinus officinalis | Ro | 29 |
| Artemisia arborescens | Aa | 42 |
| Salvia officinalis | So | 2 |
| Mentha pulegium | Mp | 4 |
| Othonna cheirifolia | Oc | 3 |
| Ebenus pinnata | Ep | 39 |
| Artemisia compestris | Ac | 8 |
| Atractylis humilis | Ah | 6 |
| Artemisia herba alba | A-h-a | 6 |
| Thymelea tartonaria | Tt | 4 |
| Total number of speicies | | 534 |

- **Biodiversity indices**

Calculated from the floristic lists established, we use several indicators of biodiversity, frequently used by biologists [9], [10].

Species richness; which represents the number of species.

The Shannon index H ' (Shannon, 1948)

$$H' = -\Sigma (p_i \log p_i)$$

Where **pi**, is the proportional abundance of species (**i**) in the sample.

$$P_i = n_i / n_j \Sigma$$

Where **ni**, is the number of individuals of species (**i**).

Its value depends on the sample size, but has the advantage of representing a number of the specific structure of the sample. It is even higher than the number of species which is large. The equitability (R), R = H / Hmax, measures the regularity of the distribution of species. The disadvantage of this index is that it presents a range of weak variation, in spit of its synthetic value. We do not retain in the study, so more as it is strongly correlated with the index of Shanon.

The advantage of this indicator is that it combines the composition and the relative importance of Two groups of the same type that have the same composition but whose species are distinct proportions constitute different entities which will have different values of index.

- For a factorial analysis of data from floristic surveys are used to XLSTAT 2013 software to describe the distribution of the floristic species in the study area.

- Diversity supra specific, aspect of the biodiversity largely neglected so far [11], corresponds to the diversity of tax evaluated on a level higher than the species. This index is important since it reveals otherness that could be masked if we considered only the species richness. We consider this diversity supra -specific level of the Gender and the Family.

- The biological type indicates the duration of vegetation of the plant. For the plants which can be presented under various types, the most common form was retained.

- The requirement with respect to the moisture of the ground indicates the preferential conditions of the plant compared to the water content of the ground.

- The mode of dissemination is identified according to the morphological attribute that the species developed to disperse their seeds. This data illustrates the aptitude of the plants to disperse across the landscape and their potential to colonize new mediums.

- The crossing of the floristic data and the descriptive characteristics now authorizes the characterization and description of the vegetation illustrated in the floristic list of the statements in the two sites.

## 3 RESULTS

### 3.1 FLORISTIC BIODIVERSITY

#### 3.1.1 DESCRIPTION OF THE FOREST OF ALEPPO PINE

The inventory carried allows, taking into account the course of the species to determine the average composition of the population per hectare.

*Table 2. Variation of stand structure of Aleppo pine*

| Class of diameter (cm) | Number of stems per hectare | Volume ($m^3$) | Basal area $m^2$ |
|---|---|---|---|
| 10 | 60 | 1.380 | 0.4712 |
| 15 | 62 | 3.844 | 1.0956 |
| 20 | 53 | 6.678 | 1.6650 |
| 25 | 34 | 7.334 | 1.6690 |
| 30 | 18 | 6.066 | 1.2723 |
| 35 | 7.1 | 3.493 | 0.6831 |
| 40 | 3.0 | 2.043 | 0.3770 |
| 45 | 1.1 | 0.999 | 0.1749 |
| 50 | 0.35 | 0.411 | 0.0687 |
| 55 | 0.20 | 0.280 | 0.0475 |
| 60 et au-delà | 0.05 | 0.090 | 0.0141 |
| Total | 238.80 | 32.628 | 7.5384 |
| **Average of the tree volume** | | | 0.136 $m^3$ |
| **The average height of the tree** | | | 8.55 m |

The data analysis of table 1 shows that young feet Aleppo pine (of class with a diameter 10-25 cm) is the most abundant: their density is from 34 to 60 stems per hectare.

The old trees (of class of diameter higher than 35 cm) are slightly met: their densities are 0.05 to 0.35 stems per hectare.

The frequency distribution of the feet of Aleppo pine marks a clear variation while going from the East towards the West.

Table 3.   Variation in the frequency of Aleppo pine trees in two different sites

| Class of diameter (cm) | Number of stems per hectare | |
|---|---|---|
| | plot 1 (West) | plot 2 (East) |
| 10 | 181 | 15 |
| 15 | 135 | 20 |
| 20 | 66 | 31 |
| 25 | 38 | 28 |
| 30 | 15 | 25 |
| 35 | 5.40 | 12 |
| 40 | 1.00 | 6.20 |
| 45 | 0.66 | 3.40 |
| 50 an beyond that | 0.46 | 1.44 |
| Total | 442.52 | 142.04 |
| **Volume / ha** | 39.29 m$^3$ | 33.15 m$^3$ |
| Class of diameter (cm) | Number of stems per hectare | |
| | plot 1 (West) | plot 2 (East) |
| 10 | 181 | 15 |
| 15 | 135 | 20 |
| 20 | 66 | 31 |
| 25 | 38 | 28 |
| 30 | 15 | 25 |
| 35 | 5.40 | 12 |
| 40 | 1.00 | 6.20 |
| 45 | 0.66 | 3.40 |
| 50 an beyond that | 0.46 | 1.44 |
| **Total** | 442.52 | 142.04 |
| **Volume / ha** | 39.29 m$^3$ | 33.15 m$^3$ |

Table 2 enabled us to notice that the young feet of Aleppo pine are more frequent in plot 1 of on the west side, while the oldest feet are met on the east side forest.

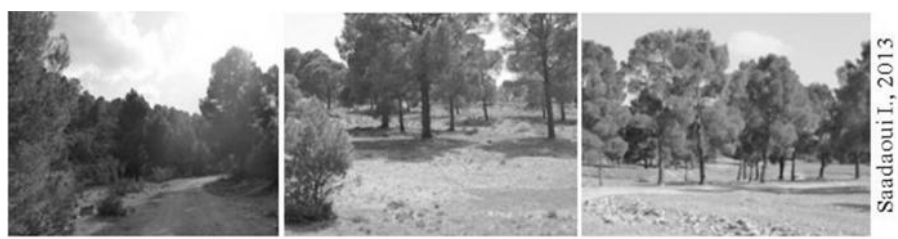

Fig. 3.   Landscape of Aleppo pine forests

## 3.2   ESTIMATION OF THE VEGETATION BIODIVERSITY OF THE SITE OF STUDY

### 3.2.1   SPECIFIC RICHNESS

Table 4.   Number of species

| | Number of species |
|---|---|
| **Site of observation of Bouchebka** | 26 |

The site of observation is of a surface of 1 hectare, which allows us to conclude that the flora of the site is rich (26 species).

### 3.2.2 FLORISTIC DIVERSITY

**- The index of Shanon, H'**

The species which represent the higher index is Stipa Caplillata with H '= 0.3539, respectively followed by shrub Globularia alypum, with H '= 0.3316, Artemisia arborescens, with H '= 0.1999, the weed Ebenus pinnata, with H' = 0.1910, the cultivated Rosmarinus officinalis which has an index H '= 0.1581, and the tree Pinus halepensis (Aleppo pine) with a index H '= 0.1392.

*Table 5.  Average of Shannon index of diversity*

|  | H'moy |
|---|---|
| Site of observation  (Bouchebka forest) | 2.4059 |

The average index of diversity of Shannon of the study area is equal to **2.406**, given that the Shannon index is between 1 and 5 (1 <H '<5), hence we note that the floristic biodiversity of this site is average.

### 3.2.3 SUPRA-SPECIFIC DIVERSITY

*Table 6.  Number of Families and Gender*

|  | Families | Gender |
|---|---|---|
| Site of observation ( Bouchebka forest) | 12 | 17 |

**- Diversity of Families**

The diversity of families in the current flora is equal to 12 (in a space of 1 hectare). It is mainly composed by the family Asteraceae, Lamiaceae, Brassicaceae, Cupressaceae and Poaceae. The total flora is dominated by the Asteraceae (presented by 8 species).

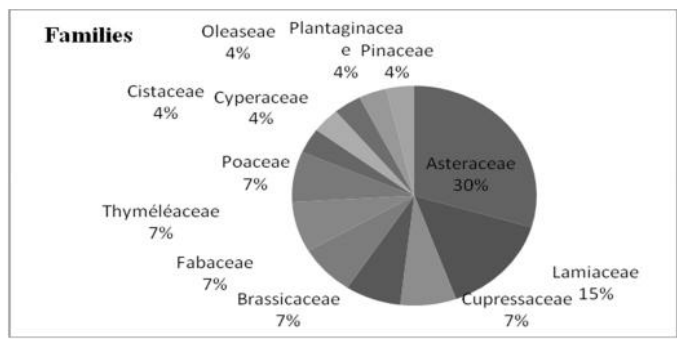

*Fig. 4.  Importance of the Families of flora in the site of observation*

**- Variety of Genders**

The full total number of the current flora is equal to 17. Most genders are represented by only one species. Artemisia is the met the most with a percentage of 16%.

In the zone of observation, we met 5 genders represented by more than one plant species. The remains of the found genders are represented by only one species.

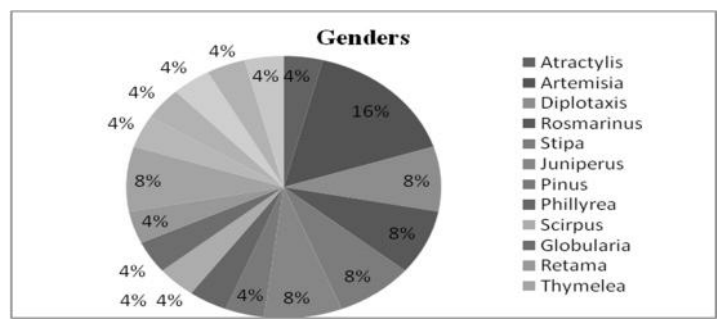

*Fig. 5.    Diversity of floristic gender in the study area*

### 3.3 DESCRIPTION OF THE VEGETATION

The description of the vegetation in our study is based on the analysis of the distribution of species according to their features of life. For that, the data of the autoecology, collected from bibliographical sources are used.

For the floristic list made up following the statements on ground, the data of autecology inform on the reasons for the distribution of the species according to their intrinsic characteristics (Figure 6).

- **Ecological Group**: vegetation is especially represented by species whose optimal habitat corresponds to forest areas with 63% (adventitious, and forest formations). The species of the thin meadows account for 23% of the totality of the floristic list.

- **Requirements of the species with respect to the light**: A majority of the species pushes under strong conditions of light (85%), which testifies to the wealth of non wooded spaces, but also not very dense wooded mediums, in connection with the strong proportion of representatives of the ecological group of the forest species. The forest species, moreover, can colonize grassy formations. This singularity can mark the installation of the progressive succession in the grassy formations with spontaneous vegetation characterized by non xerophilous forest species

- **Requirements of the species with respect to the moisture of the soil:** A majority of the species request moisture of the relatively weak soil from very weak (steppe vegetation). Few species are with broad amplitude. Hygrophilous species were not met, which is linked to the scarcity of the aquatic environments. The strong proportion of species with trend mesophilous brand of the mediums with dry trend (62% of the species met), which prevails because of a broken relief, favourable to the streaming of water.

- **Dissemination Mode:** the flora of Bouchebka is especially made up of species without specialized dissemination mode (barochore: 30%), or being disseminated by effect of the wind (anemochores : 35%), a weak distribution of the species with endozoochory (15%) and epizoochory dissemination (20%).

- **Biological type:** flora is especially represented by annual plants and perennial herbaceous plants. The woody plants (trees, shrubs, chamaephytes) represent 67 % of the flora. The strong proportion of annual and perennial herbaceous species reflects a territory dominated by open grassy spaces.

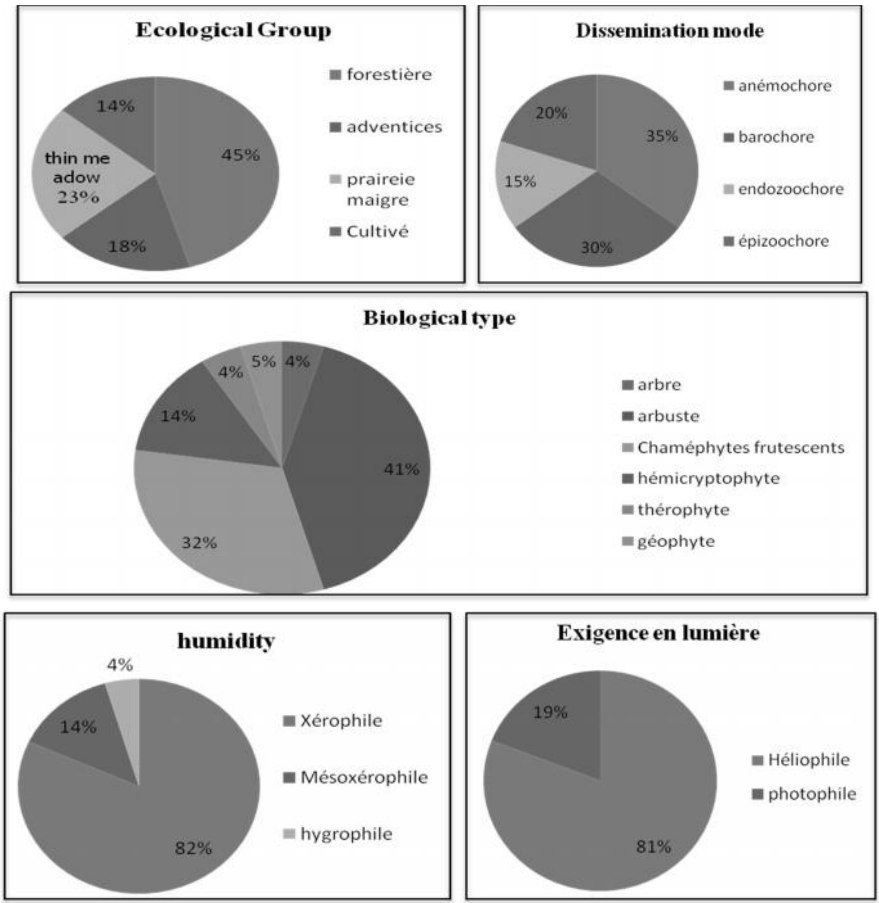

*Fig. 6. Ecological characteristics or features of life of the species of floristic list-Distribution terms for each variable considered*

Graph: factorial distribution of ecological groups in the site of study shows the projections in the factorial plane F1 x F2 profile associated with the terms of the plant species found in the study site (Figure 7).

From this graph we can draw the following table:

*Table 7. Principal coordinate of ecological groups in the factorial design*

|  | F1 | F2 |
|---|---|---|
| Forestry | 0,468 | 0,214 |
| Adventitious | -0,051 | -0,304 |
| Meadow steppe and lean | -0,362 | 0,331 |
| Cultivated | -0,008 | -0,073 |

Table 8 shows that only forest species, the steppe species and lean meadow for the factorial plane F1 x F2 present a positive index that reflects the forest and pastoral vocation midium, cultivation and crop species have a very low negative index. This shows us the low distribution of these species.

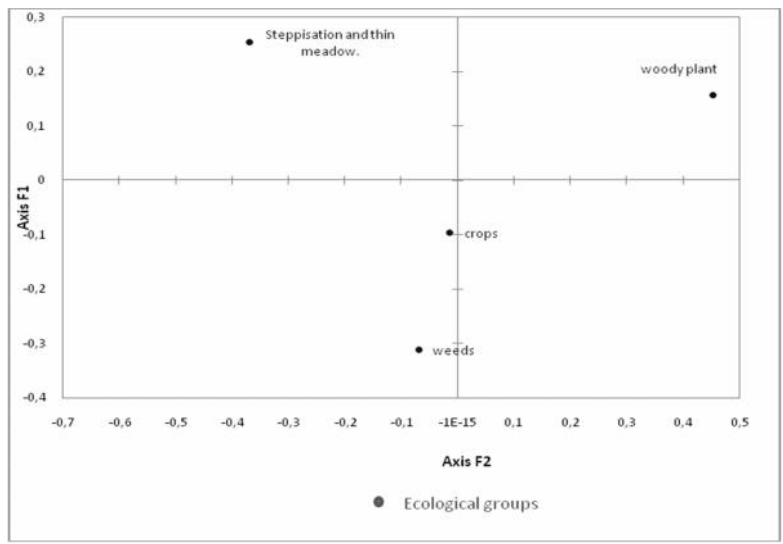

*Fig. 7.   Factorial distribution of ecological groups in the site of study*

The joint analysis of axes F1 and F2 of the factorial plan allows us to characterize the vegetation, its distribution and its specific composition, according to the feature of life of the species.

Figure 8 shows that the level of plan F1- F2 , we find on the positive side tax them with the species which presents a steppe aspect presented mainly by (Stipa tenacissima and Stipa cappillata), there is a positive trend towards the steppe and training for poor grassland pastoral trend site. We also note the presence of woody species (Pinus halepensis , Juniperus oxycedrus , Phillyrea angustifolia , Cistus creticus , Scirpus sylvaticus , etc. . ) .

On the negative side of the plan, we note the presence of crops appearance (Artemisia vulgaris, Rosmarinus officinalis, Othonna cheirifolia) marks a presence of human activities; these species have a very low indicating parameter, both the presence of weeds that accompanies the presence of crops.

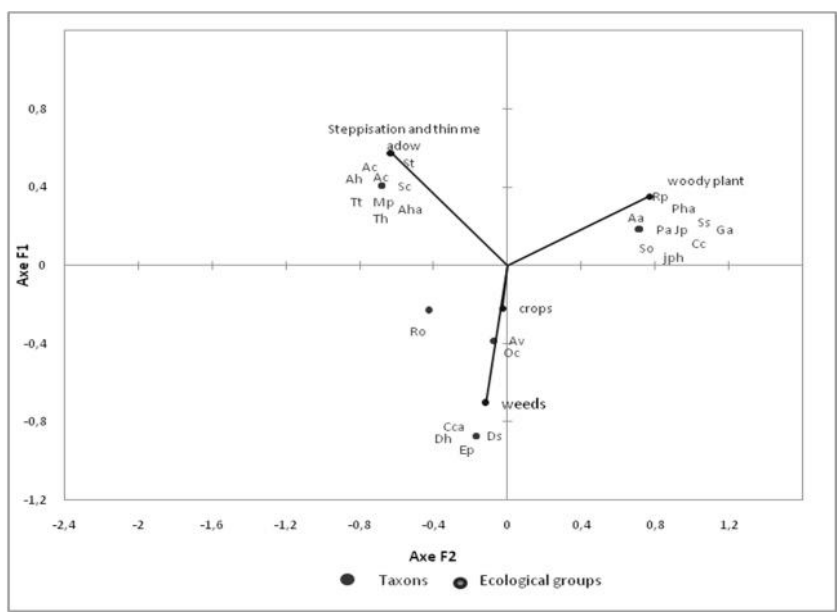

*Fig. 8.   Factorial plan showing the interaction of the species according to their ecological groups in the study site*

**Interpretation of the axis F1:**

- Negative coordinates: cultivated vegetation and weeds, conducive to the presence of human action.

- Positive coordinates: of grassland plants colonized by steppe species (Stipa Stipa capillata and tinsacissima) and a large presence of woody species.

**Interpretation of the axis F2:**

- Positive coordinates: woody plants, a practice influenced by human nature (correlated with the steppe and grassland vocation)

- Negative coordinates: grassland plants, crops and weeds marking the pastoral vocation of the area.

The vegetation is marked mainly by species revealing conditions of the dry medium mainly: considerable proportion of species to xerophilous trend, related to a rugged relief and the small proportion of the demanding water species : hygrophile trend 4% testifies weak surface to the mediums with which they are dependent (stretches water, ditches, wadis etc.). The strong proportion of species of light, annual and herbaceous long-lived testifies to a territory dominated by open and grassy spaces or not very dense wooded plantations. That make us notes that a strong proportion of species being disseminated fixed on the animals.

## 4 DISCUSSION

The results of analyzes carried on the floristic indicator and the composition of the vegetation, allow us to clarify the components of the floristic biodiversity in the zone of study. The richness of the results obtained by an analytical study thrust has enabled us to identify the importance of forms of degradation of biodiversity including human activity (the pasture, etc...). The clear presence of the young feet of Aleppo pine is the indication of stand regeneration of Aleppo pine in the western part of the forest and attenuation of the anthropogenic action toward the East, where it may notice a gradual disappearance of the old stand of Aleppo pine, this aspect is probably due to human actions (deforestation) and fires.

The floristic statements showed the difficult natural conditions of the zone according to their features of life:

The index of Shannon indicates that most of the share of the species are in fact far from abundant in the site of observation consideringthe natural conditions of the medium (scarcity ofthe factor water, nature of the grounds, an insufficient regeneration of the forest). It is an environment that is characterized by its aridity and windy side (82 % of species are drought-resistant and 36 % of species are disseminated by anemochory ) , the index of Shannon which has an average value though has made our statement in a protected area which shows the fragility of floristic biodiversity factor in this zone threatened especially by anthropisation.

The factor analyzis showed the presence of a clear pastoral steppe and appearance, as well as the anthropic actions which influence negatively the forest aspect.

All these indices reflect the fragility of the medium. Such a study will be necessary at this time, because it is now clear that the conservation of biodiversity is a priority to implement for all human activities measurement processes, for its maintenance are still far from being completely understood.

The vegetation used to evaluate the vegetal biodiversity and to reveal the impacts of the human activities on the vegetable biodiversity in the landscapes natural, such as those mountains of Bouchebka, proved to be an original and appropriate biological indicator, reproducible other landscapes. This indicator is important to develop 'landscape durability' [12].

## 5 CONCLUSION

The main rationale of our work was to define the floristic spectrum in the region of the Tunisian mid-west in general and those of the common Bouchebka in particular.

As a conclusion, this study of the flora allows the development of a floristic catalogue. The highlighting of all the richness and floristic diversity. Thus, the series of the forest area of Bouchebka are strongly influenced by human action which is very old as well as demonstrating the many Roman ruins, some of which are located in the forest where the Aleppo pine in some recolonise. Harm military occupations (campaign 1942/1943 and the Algerian conflict) is added significantly to those committed by users. Forest Bouchebka represents diverse populations resulting from the action of fire, consumers, military occupations and forest.

# Contribution to the study of the flora in the central-west of Tunisia: landscape dynamics and evaluation of plant biodiversity of mountain Bouchebka

The study of landscape dynamics in the region of Bouchebka starting from the biodiversity and the anthropic action applied to this territory shows its rich landscape scale biodiversity, but also the pressure exerted by the natural action and human threatening components of the landscape and their consistencies.

To ensure the landscape durability of these mediums, the conservation of these fragile zones from now on is regarded as being an urgent measure to be implemented for the national strategies of the development transborder.

REFERENCES

[1] N. Myers , Mittermeir CG et al. 'Biodiversity hotspots for conservation Priorities'. *Nature*, 403: 853-858, 2000.
[2] JC Matthew , Mark and TB DS Keith. Estimating the costs of soil erosion envirnmental at multiple scales in Kenyia using emergy synthesis. *Agriculture, Ecosystems and Environment*. 2006.
[3] A Kalpana. H Syed Ainul. And B Ruchi. Social and Economic considerations in conserving wetlands of Indo- Gangetic plains : A case study of Kabartal wetland , India . *Envirnmentalist* 27 : 261-273 . 2007.
[4] P Donadieu and H Rejeb. *Landscaping Chronicles of two shores of the Mediterranean*. Item. Official print editions, Tunis. p191, 2011.
[5] B Seibert and S Gharbi. *Second Forestry Development Project . Technical assistance component , Management Plan Chaambi National Park : Nature Reserve khechem el kelb Nature Reserve tella and buffer zone / boundary* . General Directorate of Forestry / Ministry of Agriculture . 80 p. 2001.
[6] CRDA Kasserine / Department of Forests. Report the forests of Feriana 1 series. p85. 1995
[7] AEA District of Research and Development / CRDA Kasserine. *State Forests Report Feriana trial Verbal From 1957 to 1995 planning*. Rapport. p89. 1995.
[8] P Quézel and S. Santa. *New flora of Algeria and southern desert regions*. Volume I, CNRS. Paris (France), pp: 201-203 . 1962.
[9] R Barbault. An ecological approach to biodiversity. *Nature, Science, Societies*, 1 (4): 322-329. 1993.
[10] A Bornard. P. Cozic and C Brau- Nogue. Species diversity of vegetation in pasture . Influence of environmental conditions and practices. *Ecology*, 27 ( 2): 103-115. 1996.
[11] M Lamotte. About biodiversity. *Mail environment INRA*, 24: 5-12. 1995.
[12]
[13] MG Paoletti. Using bioindicators based on biodiversity to asses landscape sustainability. *Agriculture, Ecosystems and Environment* , 74 : 1-18. 1999.
[14] I Saadaoui. *Landscape dynamics of natural areas Bouchebka . A test taming of mountain landscapes*. Memory Masters. Superior Institute of Agronomy - Chott Mariem, University of Sousse. 150 p. 2013.